\documentclass[graybox]{svmult}

\usepackage{mathptmx}       
\usepackage{helvet}         
\usepackage{courier}        
\usepackage{type1cm}        
\usepackage{mathrsfs}
\usepackage{amssymb}
\usepackage{amsmath}
\usepackage{makeidx}         
\usepackage{graphicx}        
\usepackage{multicol}        
\usepackage[bottom]{footmisc}

\newcommand{\kets}[1]{\left\vert #1 \right\rangle}
\newcommand{\bras}[1]{\left\langle #1 \right\vert}
\newcommand{\beq}{\begin{equation}}
\newcommand{\eeq}{\end{equation}}
\newcommand{\bqr}{\begin{eqnarray}}
\newcommand{\eqr}{\end{eqnarray}}

\makeindex             

\begin{document}

\title*{Modelling non-paradoxical loss of information in black hole evaporation}
\author{Sujoy K. Modak and Daniel Sudarsky}
\institute{Sujoy K. Modak \at KEK Theory Center, High Energy Accelerator Research Organization (KEK),\\ Tsukuba, Ibaraki 305-0801, Japan.\\ \& \\ Facultad de Ciencias -- CUICBAS,  Universidad de Colima, C.P. 28045, Colima, M\'exico\\ \email{sujoy@post.kek.jp}
\and Daniel Sudarsky \at Instituto de Ciencias Nucleares, Universidad Nacional Aut\'onoma de M\'exico, M\'exico D.F. 04510, M\'exico \email{sudarsky@nucleares.unam.mx}}
%
%
\maketitle

\abstract{We give general overview of a novel approach, recently developed by us, to address the issue black hole information paradox. This alternative viewpoint  is based on  theories involving  modifications of   standard  quantum theory, known as  ``spontaneous dynamical state reduction''  or ``wave-function collapse models'' which were  historically developed to overcome the notorious  foundational problems of quantum mechanics known  as the ``measurement problem''.   We show that   these   proposals, when   appropriately adapted and refined for this context, provide a self-consistent picture where loss of information in the evaporation of black holes is no longer paradoxical.}

\section{Introduction}
\label{sec:1}
 The black hole information problem \cite{hawking} is one of the most debated and controversial problems of theoretical physics,   and has been the focus  of considerable attention from various theoretical viewpoints during  the  last four decades (see \cite{Mathur1} for a pedagogic introduction). We  in fact  note Paddy's recent proposal \cite{Lochan1}, \cite{Lochan2} connected to   this issue. The main  reason behind this activity is the  fact that   while 
  according to  the \emph{unitary}    evolution  law of Quantum Mechanics (QM),  all information  about  a quantum  state at any time  is  encoded  in   the state  at any other time,   the process of   thermal black hole evaporation   via   Hawing  radiation poses a ``threat" to such an expectation,  leading to the so called ``paradox''.

\begin{figure}[t]
\centering
\includegraphics[scale=.4]{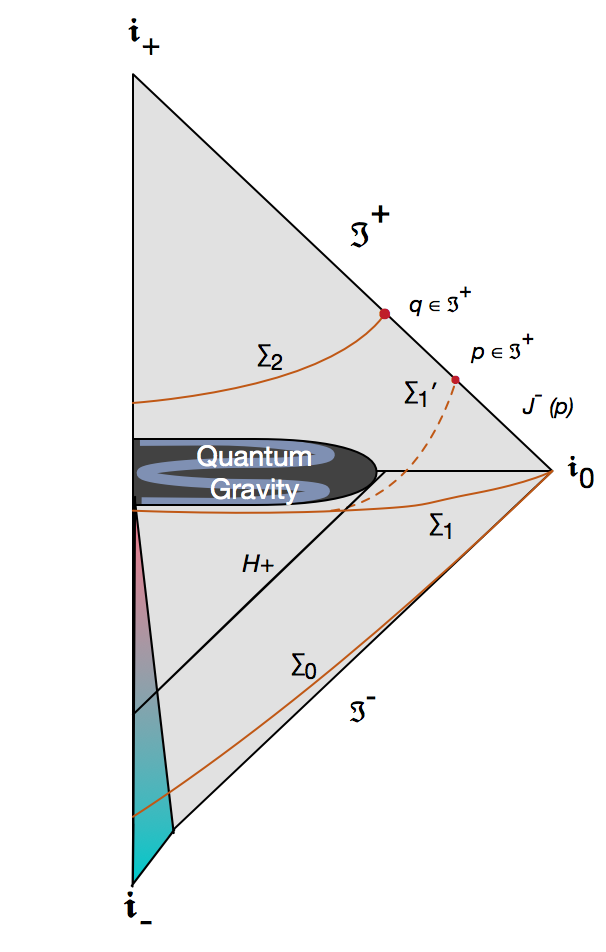}
%
%
\caption{Penrose diagram showing black hole formation and evaporation. The initial spacetime is Minkowskian at ${\cal I}^-$ and at the end of Hawking evaporation Quantum Gravity (QG) resolves the would be classical singularity making the final spacetime asymptotically flat at ${\cal I}^+$.}
\label{fig:1}       
\end{figure}

 More  specifically,  let us consider the formation (by gravitational collapse)  of a black hole and  its  subsequent  evaporation (by Hawking effect) as shown in Fig. \ref{fig:1}.  Let the initial state of the matter, defined on an initial Cauchy slice $\Sigma_0$, be characterized  at the  quantum level, by some pure,  perhaps  a coherent state.  Under appropriate   circumstances this matter  collapses forming the black hole, and any quantum field, in this  space-time   will contribute  to  the Hawking radiation  at late times. The radiation is  characterized,  in full,   simply by the   temperature  and in particular  it  will be  the same  for any  initial mass  regardless  the details of the  initial  quantum state. Thus unless some   dramatic departure  from the above  picture  takes place  there  would  be no  way to retro-dict the  initial state from the former. The mapping from initial to final states would  not be invertible and  in particular   it  would  fail to  be represented in terms of a unitary operator.  
 
 The point is  that,   attempts to explain how   the   full state of the quantum fields  \emph{unitarily}  related  to  the initial state    would  be  encoded  on late time hypersurfces   such as   $\Sigma_2 \cup  (J^-(q)  \cap {\cal I}^+$), have been  \emph{unsuccessful} until this  date.

 This, in turn, leads  the  majority of the physicists  to  believe that the fate of information and problem of unitarity in black hole evaporation is a \emph{unique situation}  where   the unitarity of quantum  evolution  is  in question . The recent finding of ``firewall problem'' \cite{Firewalls}  when considering  a solution  to the  problem based on   an approach known as  based on   `` black hole  complementarity" \cite{Susskind:1993if}  is  often  presented  as   reflecting  the  tension between the \emph{unitarity} of QM and \emph{equivalence principle} of general relativity (GR).

   We, however, need to  be a bit  more careful because  the  mapping implied  by quantum theory is  expected to  be unitary only  if it  refers  to states defined on  complete Cauchy hypersurfaces.  The point is that, while the  initial state  was characterized on   $\Sigma_0$,   which  is  a  true Cauchy  hypersurface,  the late time  hypersurfaces    would  fail to be Cauchy hypersurfaces   due to the presence of a space-time   singularity   deep within the black hole.     This  singularity  (or more precisely  a  surface   arbitrarily close to it)  can in  fact  be considered  as a  boundary of the space-time  and  thus   as a   necessary component   of any  true   late  time  Cauchy  hypersurface. However  the  singularity is   generally   expected to  be   a  feature of the  fact that  we have  not  incorporated   the aspects of  quantum gravity    which   should  cure this  singularity removing the need to   include  the extra   space-time boundary.  It is   then and  only then that we face  a truly  problematic    situation \cite{Okon-New}. 
 
 Therefore  the  real  tension  between  quantum theory  and   general  relativity 
 in the context  of  black hole  formation and   evaporation  arises   only  when  we view that quantum gravity  will remove the singularity and   thus    the need to include  a spacetime boundary.  It is  then and only then that   we face  something that could be    considered  truly paradoxical.  We  could   now ask ourselves,  what  would  be   the problem  of  adopting   the position that, all processes   involving  black hole   evaporation   do in  fact  break  the   unitarity of   quantum  
evolution? As we  see it, the problem   with that position would be that,  as  we  just   saw,  we  would  be  working  in a  context  where  we   imagine   having  incorporated    aspects of quantum gravity in the discussion.  Having  done  that, it seems inevitable to  view  processes  such as   black hole  formation and   evaporation  as   part  of a  larger  class of   processes,  after all,   the   black hole  concept  is   essentially  a  global   one. That is    the  notion of  Black Hole  is not one that  could be   considered  as lying  at the basic  formulation   of the theory,  which is  expected to be  described  in terms of  some  fundamentally local  degrees of freedom, rather  that  the global  notions  such as  event horizons,  or   trapped  surfaces  that should appear  only as  secondary and  emerging  entities.  In fact   we  should  expect that   black hole  creation and  evaporation should appear  in the theory  occurring also as  virtual  process  
contributing to  essentially all physical processes,  thus  raising the question of  when  precisely can  we  expect to  have an  exact   unitary evolution  law as  dictated  by standard quantum theory.

 In fact the  black hole information  issue,  motivated the  analysis  in \cite{Peskin}    where  it  was  argued that   loss   of unitary   would be need to  accompanied    by  unacceptably large violations of   energy conservation or  of   causality.  A  subsequent  study of   the issue reveled  however that those   arguments  were not    very  robust, and that such  expectations  could  be  radically modified \cite{Unruh-Wald}.

  These  considerations  open the door to  considering the question of information  loss in  the context of    possible   of modified  versions of quantum theory  involving   departure from unitary evolution at the fundamental level.

%

In fact  it is fair to say that all  the approaches that  have  been  proposed   so far  for   the recovery of information  (and the  full quantum state that is  unitarily related to the initial  one) have not been successful as  they end  up  adding to more problematic  aspects to the picture. A big motivation of this rather ``one way traffic'' is the  adherence to the notion that  failure of  unitarity in black hole evaporation would    completely invalidate  QM. However, as we advocate here, the situation is  not as simple,   because the violation of unitarity is not only not  unexpected in  quantum  context    but  rather a common occurrence  in any  situation,   normally thought  as involving the collapse of wave-function due to whatever reason (natural interaction or laboratory measurement).  That is,   we  have further motivation  to  consider   the issue   at   hand, in connection with the so called   general measurement problem in quantum theory. 

Therefore,  in contrast  with the  established  tradition  we  will discuss   here  an approach   based on  the  exact opposite possibility, i.e.  the  \emph {necessity  of  loss of information} during black hole evaporation just as in  most    ordinary  situations  involving   the quantum   regime,  thus   ``dissolving'' the paradox.

  This    ``dissolution" comes  of   course  at the cost of losing  quantum mechanical unitarity  and  one  might worry  whether  we  would  lose  with it   all the successes of  standard QM.
 We  furthermore note  that  from the foundational   point of  view regarding  quantum theory, there  have been   various  proposal of a modified version of quantum dynamics incorporating  a \emph{spontaneous collapse} of the wave-function to elevate QM from a \emph{theory of measurement} to the \emph{theory of reality} (see \cite{newcsl} for the terminology) in a manner that the subjective role of an observer becomes removed and one can treat QM  objectively without introducing any  extraneous  notion of observer  as an essential  entity  shaping reality. This collapse process is spontaneous and stochastic, and it is implemented in such a manner that the  well established  and  experimentally  successful predictions of quantum mechanics remain unaffected, while a gradual difference in the predictions  appears as the  quantum system's size approaches  that  of macroscopic  object (for an account of ongoing experimental endeavor, see \cite{Bassi} ).   Thus, in  building such type of  theories, the aim is     to resolve \emph{the measurement problem} and eliminate various in-built two level descriptions of reality in Copenhagen interpretation; such as, micro/macro, classical/quantum, system/apparatus, system/observer, system/environment etc. 

In this article, we will   not discuss in detail any of these proposals (for that we refer the reader to the papers \cite{GRW:85}-\cite{rel2}, \cite{Pearle:76}-\cite{newcsl}, \cite{Bedingam-Rel, rel11, RelPearle, Tumulka-Rel, Tumulka-Rel-1} as well as review articles \cite{Bassi:2003gd, Bassi}), but we shall use some specific models and provide a concrete example  offering a  overview of the manner in  which such models  can deal  with  the information problem in black holes leading to  a picture   where   the  associated    breakdown of unitarity   is a part and  parcel of  the  (modified) general  quantum mechanical  evolution.

\section{Measurement problem and models of wave-function collapse}
\label{sec:2}
According to the Copenhagen interpretation of quantum mechanics there are two distinct evolution  rules for the  quantum state/wave-function  of a  system.  First, a continuous evolution as dictated by the Schr\"odinger equation and valid while  the  system is  left  alone  and free from observations,   and  the  second,  a discontinuous and stochastic jump to one of the eigenstates (dictated by Born probability rule) of some self-adjoint operator in the Hilbert space once \emph{measurement}  by an external observer  takes place. As characterized by R. Penrose, the first case is a unitary evolution or the $U-$process, while the second case is a reduction or the $R-$process, and \emph{measurement} is a notion that separates these two processes.  The problem is that within the standard view of QM, measurement does not have any  kind of rigorous  definition,  nor   is it clear when exactly it is performed during a evolution. In fact such   vague and artificial  division has  been  sharply  criticized by J. Bell \cite{JBell}
\begin{quotation}
...If the theory is to apply to anything but highly idealized laboratory operations, are we not obliged to admit that more or less `measurement-like' processes are going on more or less all the time, more or less everywhere? Do we not have jumping then all the time?
\end{quotation}
Also, it is an in-built aspect of  the  standard  presentations, that without observer or some entity which is ``measuring'' the system,  no specific outcome is  presupposed in QM. 

Within the community, working on the foundation of quantum mechanics, there are of course    diverse  viewpoints regarding  the \emph{measurement problem}. These  include  the Many  World  Interpretations,  in  its  various   forms,  the  Bohmian Mechanics   program  representing a reliance on nonlocal hidden variables,  and   the  proposals for   unifying the $U$ and $R$ processes, referred as the Dynamical Reduction Program (DRP), pioneered by Pearle, Ghirardi, Rimini and Weber. 

The DRP first included a discrete process of collapse in the wave-function/quantum state, driven by an additional non-unitary  and  stochastic  term   modifying  the Schr\"odinger evolution.

 The   basic idea  in those proposals  is that  the evolution  of   systems   with very   small number  of  degrees of freedom   is   dominated   by the standard  part of   the dynamics resulting   in very small  deviations      from  that predicted  by standard  theory,  ensuring  the reproduction of  the  stupendous success  of   quantum theory     in  high precision   laboratory  experiments 
whereas, the non-standard  terms  becomes  dominant   when  a   rather large  number of degrees of freedom  appear in a  state   representing  a  rather  delocalized  quantum  superposition, thus  ensuring the  rapid  collapse  to  one or the other  of the  classical looking  components of   Schr\"odinger cat states.

  This  feature  ensures that   when,  what is normally called  a measurement is performed, the   system is  driven to one or the other eigenstates of  the  apparatus'   pointer's   position    simply  because such pointer   consists of  a  macroscopically large  number of    degrees of freedom.   That is,   as   a result of the  new  general  dynamical  law  the theory reproduces the   standard  predictions of quantum theory regarding the measurement    of  \emph{the appropriate   self-adjoint operator}. The   first  and is   simplest  successful  model of this  kind  known as Ghirardi-Rimini-Weber (GRW) theory \cite{GRW:86}, which was later improved to  deal with   identical particles  in a scheme  that  makes wave-function collapse a continuous process and  known as the CSL theory \cite{GRW:90, moreCSL, newcsl}. Their  recent  advances  in this  direction have resulted   in  proposals  for     relativistic  version of  both type of theories \cite{Bedingam-Rel, rel11, RelPearle, Tumulka-Rel}. In this article we restrict ourselves to the non-relativistic framework of CSL theory to adress the issue of black hole information and for the relativitic framework we refer the interested reader to our recent work \cite{BMS}.

\subsection{CSL theory: non-relativistic setting}
\label{subsec:2}
The non-relativistic version of the CSL theory \cite{GRW:90, moreCSL, newcsl} is, at this point, much  better  explored than the relativistic counterpart and it is defined by following two equations: 
(i) A  stochastically modified Schr\"odinger equation, whose  solution is:
  \begin{equation}\label{CSL1}
 { |\psi,t\rangle_w = \hat {\cal T}e^{-\int_{0}^{t}dt'\big[i\hat H+\frac{1}{4\lambda_{0}}[w(t')-2\lambda_{0}\hat A]^{2}\big]}|\psi,0\rangle,}
\end{equation}
 where $\hat {\cal T}$ is the time-ordering operator,  $w(t)$ is a random, white noise type classical function of time and its probability distribution is given by the second equation,  (ii) the Probability Distribution (PD) rule:
  \begin{equation}\label{CSL2}
	 { PDw(t)\equiv{}_w \langle\psi,t|\psi,t\rangle_w \prod_{t_{i}=0}^{t}\frac{dw(t_{i})}{\sqrt{ 2\pi\lambda_{0}/dt}}}.
\end{equation}
Thus  the  standard   Schr\"odinger    evolution   and corresponding changes in the state corresponding  to a ``measurement" of  the operator {$\hat A$} are  unified and the dynamics does not allow any superluminal signaling. In the  non-relativistic limit, for a single particle,  the proposal assumes that  there is a    spontaneous and continuous  reduction  characterized  by $\hat A = \hat {\vec X}_\delta $, where  $\hat{\vec X}_\delta$ is  a suitably  smeared position operator  (with the  smearing characterized by  the scale $\delta$). This smearing of the position operator is required to avoid an uncontrolled increase in energy associated with a point-like collapse event. The resultant theory can be applied to all situations without invoking any  measurement device  or observer. This framework can be easily extended to multiparticle system by choosing a set of operators representing each particle so that  everything,  including, the  apparatuses are treated  quantum mechanically. The final theory, thus  seems  to   successfully address the   measurement problem and completely overlook various two level descriptions in Copenhagen interpretation.  

Since the final outcome of the collapse of an individual state vector is uncertain, it is useful to consider a collection of identical initial state and describe the evolution of an ensemble in the language of a density matrix. The CSL evolution of density matrix can be derived from a Lindblad type equation with a solution \cite{GRW:90, moreCSL, newcsl}
\begin{equation}
\rho (t) = \hat {\cal T} e^{-\int_{0}^{t}dt'\big[i(\underrightarrow{\hat H}-\underleftarrow{\hat H}]+\frac{\lambda}{2}[\underrightarrow{\hat A}-\underleftarrow{\hat A}]^{2}\big]}\rho (0)
\end{equation}
where the arrows mean the operators act on the left or right of $\rho(0)$.

\section{Callan-Giddings-Harvey-Strominger (CGHS) Model}
\begin{figure}[h]
\centering
\includegraphics[scale=0.35]{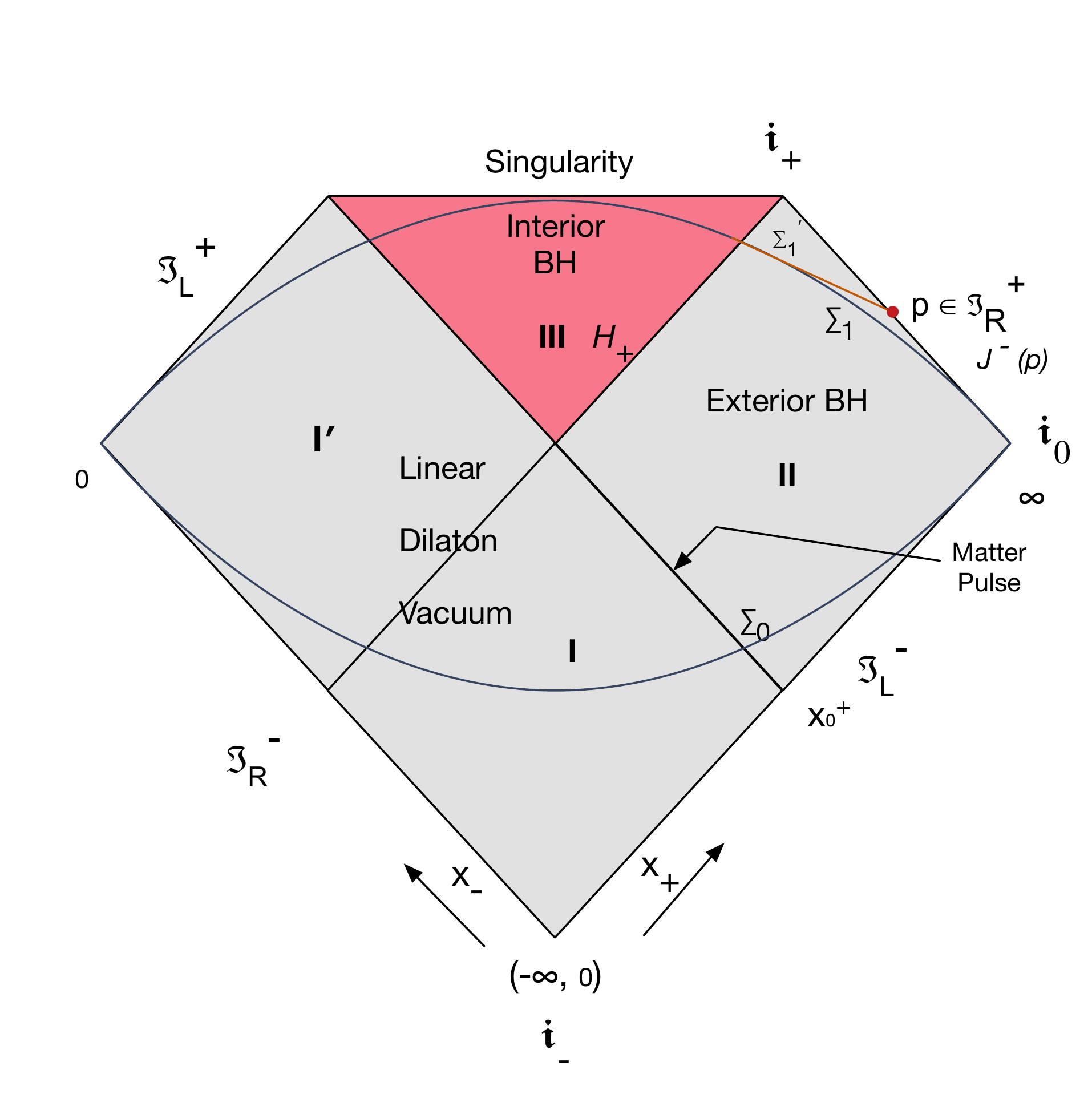}
\caption{Penrose diagram for CGHS spacetime. Minkowskian and black hole regions are separated by a sharp gravitational collapse of null matter like photon.}
\label{model}
\end{figure}

We choose the  2 dimensional version of black hole formation and evaporation,  provided by the CGHS model \cite{CGHS92}, to exhibit an explicit realization of our proposal. The CGHS action is given by
\begin{equation}
{ S=\frac{1}{2\pi}\int d^2x\sqrt{-g}\left[e^{-2\phi}\left[R+4(\nabla \phi)^2+4\Lambda^2\right]-\frac{1}{2}(\nabla f)^2\right]} \nonumber
\end{equation}
where $\phi$ is the dilaton field,  $\Lambda^2$ is a constant, and  $f$ is a  real scalar  field,  representing  matter. The Penrose diagram of CGHS model is shown in Fig. \ref{model}.
Before the gravitational collapse ($x^+ < x^+_{0}$), the metric is Minkowskian, usually known as the dilaton vacuum (region I and I'), given by ${ds^2=-\frac{dx^{+}dx^{-}}{-\Lambda^2 x^{+}x^{-}},}$ whereas, at $x^+ > x^+_0$ it is represented by the black hole metric (region II, III){\footnote{more precisely, region I', although flat, is also part of  the interior of the event horizon as nothing in that region can ever reach ${\cal I}_R^+$}} $ds^2=-\frac{dx^{+}dx^{-}}{\frac{M}{\Lambda}-\Lambda^2 x^{+}(x^{-}+\Delta)}.$ In regions I and I', natural Minkowskian coordinates are  $y^+ =\frac{1}{\Lambda} \ln(\Lambda x^+) , y^- = \frac{1}{\Lambda} \ln(-{\frac{x^-}{\Delta}})$, with $-\infty < y^{-} <\infty;~ -\infty <y^{+} < \frac{1}{\Lambda}\ln(\Lambda x_0^+)$. On the other hand, on the BH  exterior (region II), where  physical observers might  exist, one has  the coordinates $\sigma^+ = \frac{1}{\Lambda} \ln(\Lambda x^+) = y^+, \sigma^- = -\frac{1}{\Lambda} \ln(-\Lambda(x^- + \Delta))$ and the metric is $ds^2=-\frac{d\sigma^{+}d\sigma^{-}}{1+(M/\Lambda)e^{\Lambda(\sigma^{-}-\sigma^{+)}}}$ with $-\infty < \sigma^{-} < \infty$ and $\sigma^{+} > \sigma_{0}^{+}= \frac{1}{\Lambda} \ln(\Lambda x_0^+).$ It is easy to check the asymptotic flatness of the black hole metric by introducing Schwarzschild like time $t$ and space $r$ coordinates \cite{mops1,mops2} using $\tanh(\Lambda t) = T/X$ and $-\frac{1}{\Lambda^2}(e^{2\Lambda r} - M/\Lambda) = T^2-X^2$.

 The  quantum description of the field  $f$ can be made using  two different  natural  bases.  In the asymptotic  past  (${\cal{I}}^{-}_{L} \cup {\cal{I}}^{-}_{R}$ or {\it in}) region, the basis mode functions are chosen  to be:
$
{u_{\omega}^{R}=\frac{1}{\sqrt{2\omega}}e^{-i\omega y^{-}}}
$
and
$
{u_{\omega}^{L}=\frac{1}{\sqrt{2\omega}}e^{-i\omega y^{+}}},
$
with {$\omega>0$} ({$R$} and {$L$} indicate right and left moving modes respectively).
The tensor product of respective vacuum state defines the {\it in} vacuum ({$\kets{0_{in}}_{R}\otimes \kets{0_{in}}_{L}$}). In the asymptotic future (out region)  we use  a basis of  modes  that   have support in the  outside (exterior) and inside (interior) to the event horizon. The   mode  functions  in the exterior to the  horizon are:
$
{v_{\omega}^{R}=\frac{1}{\sqrt{2\omega}}e^{-i\omega\sigma^{-}}\Theta(-(x^{-} + \Delta))}$
and
$
{v_{\omega}^{L}=\frac{1}{\sqrt{2\omega}}e^{-i\omega\sigma^{+}}\Theta(x^{+} - x_0^{+}).}$
Similarly,  and in order to have a  complete  {\it out} basis,  one  chooses   a set of modes  for  the black hole interior. Usually for the left moving modes,  one maintains the same   functional form as before, and for  the right moving modes  one takes:
$
{\hat{v}_{\tilde{\omega}}^{R}=\frac{1}{\sqrt{2\tilde{\omega}}}e^{i\tilde{\omega}\sigma_{in}^{-}}
\Theta(x^{-} + \Delta)}
$.
  
It is  convenient  to replace the above delocalized plane wave modes by a complete orthonormal set of discrete and  sharply localized  wave packets modes \cite{cghs-2, cghs-3}, 
  \beq
v_{nj}^{L/R}=\frac{1}{\sqrt{\epsilon}}\int_{j\epsilon}^{(j+1)\epsilon}d\omega e^{2\pi i\omega n/\epsilon}v_{\omega}^{L/R},
\eeq
where the integers $j\ge 0$ and  $-\infty <n<\infty$. These wave packets are peaked about $\sigma^{+/-}= 2\pi n/\epsilon$ with width $2\pi/\epsilon$  respectively.

   The non-trivial Bogolyubov transformations are only relevant   in the right moving sector, and  are   the ones that  in fact account  for the Hawking radiation.  The initial   state,   corresponding  to  the   vacuum   for the right moving modes, and the    left moving pulse   (which  leads to the  formation of  the black hole)  $\kets{\Psi_{in}} = \kets{0_{in}}_{R} \otimes \kets{Pulse}_{L}$   can be  expanded in the \emph{out} basis:
\begin{equation}
N \displaystyle\sum_{F_{nj}} C_{F_{nj}} \kets{F_{nj}}^{ext} \otimes \kets{F_{nj}}^{int}\otimes \kets{Pulse}_{L}, \label{inst}
\end{equation}
where the states $\kets{F_{nj}}$  are characterized by the   finite occupation numbers  $\lbrace F_{nj} \rbrace $ for each  corresponding mode $n, j$; {$N$} is a normalization constant, and the coefficients {$C_{F_{nj}}$}'s are determined by  the Bogolyubov transformations.

\section{Gravitational induced collapse of wave-function and loss of information}
\label{sec:3}
As we have mentioned in the very beginning,  the wave-function collapse model, as given by the CSL theory, needs to be adapted  in  order to  be applicable to the problem at hand. One novel addition to the already developed  CSL theory, is our hypothesis that \emph{gravitational field enhances the rate of wave-function collapse}. This can be achieved by making the collapse rate as a function of the Weyl curvature scalar $W_{abcd}W^{abcd}$ as first suggested by Okon and Sudarsky \cite{Okon1}. That is, even in the absence of any measuring device/observer, in a spacetime region with enormously large curvature (such as inside the horizon and towards the center of a black hole), quantum superpositions are increasingly broken in a \emph{stochastic} manner (provided by the CSL stochasticity) and produces  similar  effects  as  those  caused  by an external measurement usually   considered in  a laboratory context. It should be mentioned that such an effect of gravitation on quantum mechanics was  strongly  advocated by R. Penrose in several of his works (see, for instance, \cite{penrose1, penrose2}) and we consider those as a guiding path leading to our explicit demonstration.

We first note that in the multi-particle system, the CSL evolution  (\ref{CSL1}) is generalized to
\begin{equation}
{ |\psi,t\rangle_{w_\alpha} = \hat {\cal T}e^{-\int_{0}^{t}dt'\big[i\hat H+\frac{1}{4\lambda_{0}}\sum_{\alpha}[w_\alpha(t')-2\lambda_{0}\hat A_\alpha]^{2}\big]}|\psi,0\rangle,}
\label{CSL11}
\end{equation}
where    $\alpha$ is  an  index labeling the   set of   particles. Our aim is to consider a CSL evolution (analogous to (\ref{CSL11})  applied  to a field theory   with  the  index $\alpha $ having  a  different   meaning) of the initial state (\ref{inst}). Moreover we  will treat CSL as an interaction term, so that  the free Hamiltonian  will be   set   to  zero, i.e $H=0$ in (\ref{CSL11}). As we are using this equation in the context of QFT in curved spacetime, we need to choose a new operator, that  must  be constructed using the field operator and its derivatives. One such operator is the number operator (for right moving modes) defined in the interior Fock basis times the identity for the exterior Fock basis:
\begin{equation}
{\hat {A}^\alpha =\hat{N}^{int}_{nj} \otimes \mathbb{I}^{ext}} 
\label{csl-op}
\end{equation}
for all $n, j$, where $\hat{N}^{int}_{nj} = \hat{N}^{int (R)}_{nj} \otimes \mathbb{I}^{int(L)}$ and $\mathbb{I}^{ext}= \mathbb{I}^{ext(L)} \otimes \mathbb{I}^{ext(R)}$. This ensures that the collapse will make the wave-packet to peak about particular values of $n$ and $j$ (which will be picked randomly depending on the specific realization of the noise $w_{nj}(t)$). In the standard CSL type evolution (\ref{CSL11}) it takes,   strictly  speaking, an infinite  amount of time to  fully collapse the wave-function to an eigenstate of the collapse operator due to the finite  value of the collapse parameter $\lambda_0$. It is to be noted that the experimental bounds on $\lambda_0$ come from laboratory based experiments that are,  of course, performed in a spacetime regions where curvature is negligible.  Here we make a hypothesis that the collapse rate is in fact sensitive to the local curvature, so that a more general expression must have the following form
\begin{equation}
\lambda (W^2) = \lambda_0\left(1+ ({W^2}/{\mu^2})^{\gamma}\right)
\label{lw}
\end{equation}
where $\mu$ is an appropriate scale and $\gamma\ge 1$. One anticipated effect of this, is to  generate   an  complete effective collapse of the quantum state,  taking place  in  a finite  time interval.


In the particular case of 2D models the  Weyl scalar  vanishes   identically  so instead, for this special case, we assume  that   $\lambda $ is determined  by the Ricci scalar.  Thus, for this specific case \eqref{lw} is replaced  by,
\begin{equation}
\lambda(R) = \lambda_0\left(1+ (R/\mu)^{\gamma}\right)
\label{lr}
\end{equation}
where, for the CGHS black hole  $R = \frac{4M\Lambda}{M/\Lambda - \Lambda^2(T^2-X^2)}$. The Kruskal time and space coordinates are respectively  $T=\frac{x^+ + x^- + \Delta}{2}$ and $X=\frac{x^+ - x^- - \Delta}{2}$. Next  we provide a brief account of calculation, further details can be found in \cite{mops1}, \cite{mops2}. 
 
\begin{figure}[h]
\centering
\includegraphics[scale=0.30]{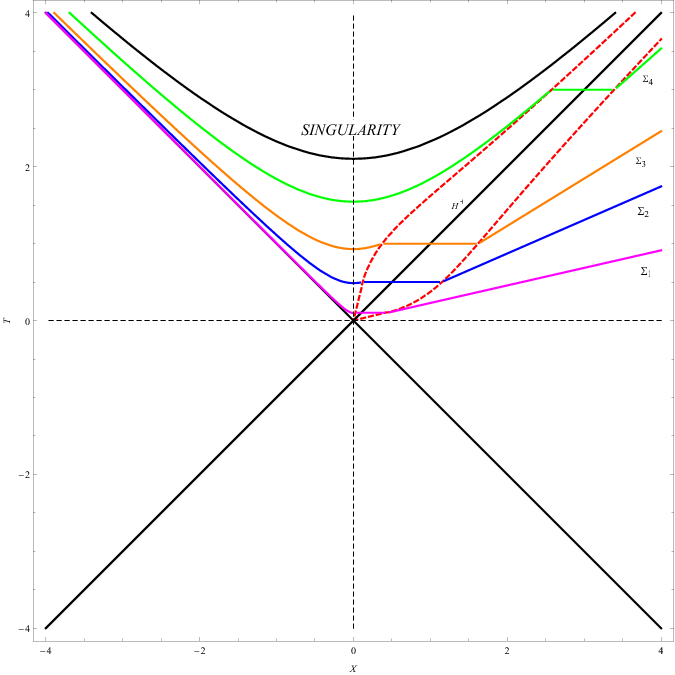}
\caption{Foliation of the CGHS spacetime by Cauchy slices. All the plots have fixed $M/\Lambda^3 = 4.42$, while $T= 0.1, 0.5, 1, 3$ (in magenta, blue, orange and green) as the joining $T=const.$ curves.}
\label{fol}
\end{figure}

{\it Foliation of the spacetime:}  To implement CSL in the  black hole model we need to foliate the spacetime with an appropriately defined spacelike/Cauchy slices, as plotted in Fig. \ref{fol}, and use the time evolution from one slide to the other.  As there is no   natural  notion  of ``time''  in the relevant regions of these  spacetimes, we   take  a convenient  time  parameter ($\tau$) that will be  used to  characterize  the   quantum  state  evolution,  according to  the  CSL  dynamics.

We define the Cauchy slices to be Schwarzschild $r=const.$ inside the horizon and Schwarzschild  $t=const.$ outside the horizon, and join them by surfaces with Kruskal $T=const.$. The intersection curves joining the family of Cauchy slices $r=const.$ and $T=const.$ at one end (inside horizon) is chosen to be $T_1(X) = \left(X^2 + \frac{M}{\Lambda^3}e^{-2\Lambda/\sqrt{X}}\right)^{1/2} $, whereas, at the other end (outside horizon) the intersection curve $T_2(X)$ can be found just by using a reflection about the event horizon $T=X$.   We fix the value  of  the ``time'' parameter $\tau$ as the value of the coordinate $T$  on  the intersection of $r=const.$ with $X=0$ (or $T$ axis), so that the Ricci scalar is expressed as $R= \frac{4M\Lambda}{M/\Lambda -\Lambda^2 \tau^2}$. It is now clear that $R$ diverges for some finite value of $\tau=\tau_s = \frac{M^{1/2}}{\Lambda^{3/2}}$  corresponding to the divergence of $R$  that characterizes the singularity.

\begin{figure}[t]
\centering
\includegraphics[scale=.2]{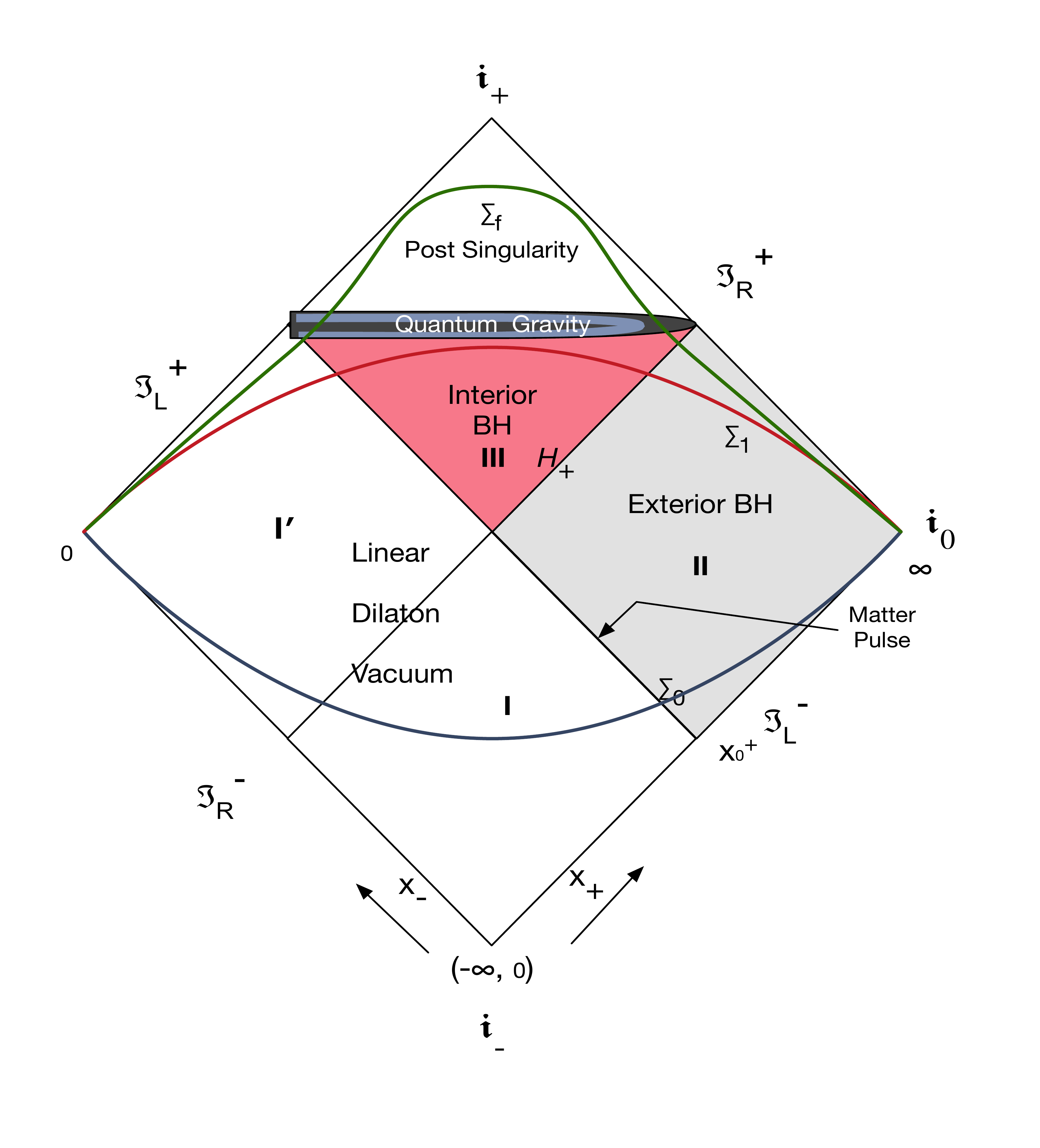}
\caption{Penrose diagram of CGHS spacetime with Quantum Gravity (QG) region.}
\label{fig:3}       
\end{figure}

{\it Evolution of the quantum state:} In standard CSL theory state is evolved according to the equation \eqref{CSL11}, which, in the present situation is subjected to the changing collapse parameter \eqref{lr}, that become a  function of time parameter $\tau$ and,   the  collapse operators \eqref{csl-op}. The initial state for the right moving modes, traveling from ${\cal I}^{-}_{R}$ to ${\cal I}^{+}_{R}$ is denoted by the ``in'' vacuum for right moving modes, which can be expressed in the ``out'' basis according to \eqref{inst} (we leave for the moment the ``pulse''  which is left moving and forms the black hole). The evolution equation of the state vector for right moving modes become
\begin{equation}
|\Psi, \tau\rangle_R = N\sum_F C_{F_{nj}} e^{-\int_o^\tau d\tau' [\frac{1}{4\lambda} \sum_{n,j}(w_{nj} - 2\lambda F_{nj})^2]} |F_{nj}\rangle^{int}_R\otimes|F_{nj}\rangle^{ext}_R
\label{stev}
\end{equation}
where $F_{nj}$ is the eigenvalue of the operator $\hat{N}_{nj}^{int}$ while acting on the state $|F\rangle^{int}$. As $\tau$ approaches $\tau_s$ the Cauchy slices tend to reach the spacetime singularity, and $R$ diverges. This divergence in $R$ makes the integral in \eqref{stev} divergent, and thus the initial state collapses to a state with definite quantum numbers $n,j$, giving
\begin{equation}
\lim_{\tau \rightarrow \tau_s} |\Psi, \tau\rangle_R = N C_{F_{n_{0} j_{0}}} |F_{n_{0}j_{0}}\rangle^{int}_R\otimes|F_{n_{0}j_{0}}\rangle^{ext}_R,
\end{equation}
on the hypersurface $\Sigma_1$ (Fig. \ref{fig:3}) as  it approaches  the  singularity. As the    level of each mode's  excitation depends on the realization of the stochastic value $w_{nj} (\tau_s)$, the final state after collapse, although remains pure, it is undetermined. 

{\it Evolution at the ensemble level as given by the density matrix:} To account for the lack of predictability of the final state we consider a large collection of systems   all prepared  in the same initial state and use an ensemble description in terms of a density matrix. The evolution equation becomes
\begin{equation}\label{rhot}
\rho_R (\tau) = N^2 \sum_{F,G} e^{-\frac{\pi}{\Lambda} (E_F +E_G )} e^{- \sum_{nj} (F_{nj}-G_{nj})^2\int_{\tau_0}^{\tau}d\tau'\frac{\lambda(\tau')}{2}} 
\kets{F}^{int}_R\otimes\kets{F}^{ext}_R \bras{G}^{int}_R \otimes\bras{G}^{ext}_R.
\end{equation}
Therefore, near the singularity (on $\Sigma_1$ in Fig. \ref{fig:3}), as $\lambda$ diverges in the exponential factor, the result is  a diagonal density matrix of the form (omitting $n,j$ from subscript for simplified notation and putting explicit expression for $C_F$):
\begin{equation}
\lim_{\tau\to\tau_s} \rho_R (\tau) = N^2 \sum_{F} e^{-\frac{2\pi}{\Lambda} E_F}  \kets{F}^{int}_R\otimes\kets{F}^{ext}_R \bras{F}^{int}_R \otimes\bras{F}^{ext}_R, 
\end{equation}
where $E_F = \sum_{nj} \omega_{nj}F_{nj}$  is the total energy of the  final  excited state.

The description of the state vector and density matrix is complete once  we include the left moving matter pulse, so that
\begin{eqnarray}
&&\lim_{\tau \rightarrow \tau_s} |\Psi, \tau\rangle_R = N e^{-\frac{\pi}{\Lambda} E_{F_0}} |F_0\rangle^{int}_R\otimes|F_0\rangle^{ext}_R\otimes |Pulse\rangle_L \\
&&\lim_{\tau\to\tau_s} \rho (\tau) =N^2 \sum_{F} e^{-\frac{2\pi}{\Lambda} E_F}  \kets{F}^{int}_R\otimes\kets{F}^{ext}_R \bras{F}^{int}_R \otimes\bras{F}^{ext}_R \otimes \kets{Pulse}_L\bras{Pulse}_L.
\label{rhot3}
\end{eqnarray}
where $F_0$ is understood as a specific  particle excited state $F_{n_{0}j_{0}}$ and $E_{F_0}$ is the energy of this state.

{\it Quantum gravity (QG) and resolution of singularity:} To pass from the hypersurface $\Sigma_1$ to $\Sigma_f$, in Fig. \ref{fig:3}, one has to   rely on a theory of QG   which is  likely to involve  giving  up the   classical notion of ``spacetime'' . In the absence of any  fully  workable theory of that kind, we make a  few  natural assumptions about QG theory, namely that   -- (i) it resolves the singularity and leads  on the other side, to  a  regime  describable with    standard classical notions of   space-time , (ii) it does not  lead to arbitrarily  large  violations of   standard   conservation  laws such a  energy conservation. If so, then QG makes following operation after combining the negative energy state $|F_0\rangle^{int}_R$ (which is complementary to Hawking radiation) with the positive matter $|Pulse\rangle$
\begin{equation}
|F_0\rangle^{int}_R \otimes |Pulse\rangle_L \rightarrow |p.s\rangle,
\end{equation}
where $|p.s\rangle$ is a post-singularity quantum state with almost vanishing energy and residing as the complement of Hawking radiation near ${\cal I}_R^+$ on $\Sigma_f$. Then on the final hyper-surface $\Sigma_f$, the quantum state and the density matrix becomes
\begin{eqnarray}
&&|\Psi \rangle_R = N e^{-\frac{\pi}{\Lambda} E_{F_0}} \otimes|F_0\rangle^{ext}_R \otimes |p.s\rangle\\
&&\rho =N^2 \sum_{F} e^{-\frac{2\pi}{\mu} E_F}  \kets{F}^{ext}_R \bras{F}^{ext}_R \otimes \kets{p.s} \bras{p.s}. \\
&&~~ = \rho^{ext}_{thermal}\otimes {\mathbb{I}^{ext}_{p.s.}}
\label{rhot3}
\end{eqnarray}
 The resulting picture,  therefore, indicates  that the final state on the Cauchy slice $\Sigma_f$, for an individual system  is {\it pure,  yet undetermined} while  at the ensemble level it is {\it   proper mixed  state} as the density matrix is  clearly thermal on the asymptotic  regime  times a  state with a very low energy  ( and idealized to be   vacuum ) characterizing the remaining  portion of  $\Sigma_f$, (which is then  taken to  be  a  portion of  flat  spacetime). Thus the complete evolution is \emph{non-unitary and information is lost}, mainly in the interior of the black hole, as a consequence of wave-function collapse. There is  of course \emph{nothing paradoxical} in this picture.


\section{Discussions}
We have put forward a novel proposal  involving  gravitational influenced wave-function collapse  which we showed can account for the enormous loss of information in black hole evaporation which  thus   leads to a  dissolution  of the so called ``paradox''. On a broader perspective, this opens up a rather interesting   possibility: that in the energy scale interpolating between,  say, the current LHC (or the Standard Model) energy scale (about 10 TeV) and somewhere below the QG scale\footnote{up to the  regime  where we  expect the  validity of   standard space-time notions  as  provided  by General Relativity.} ($10^{16}$ TeV), there  could   be  important  effects describable in the context   of  \emph{the  standard model of particle physics    adapted to the   modified  quantum field theory constructed on  curved space-times with the  additional feature of gravitational induced quantum state reduction,   as provided,  say,    by   one  of the relativistic  collapse proposals \cite{Bedingam-Rel, rel11, RelPearle, Tumulka-Rel, Tumulka-Rel-1}}.
Could it be,  for instance, that the  issue of  the  radiative  corrections  induced quantum  instability of  the Higgs potential  are  modified by the introduction of  such modifications?    Could   we  do  with a   scheme  where  supper-symmetry  is not   needed  and   have   similar benefits  arising,    instead,  from   the  effects of   quantum collapse?   We believe this line  of inquire might  offer  interesting   insights and 
modify  the perspectives for  physics beyond standard model,   and  perhaps  the  expectations   for  phenomenology of quantum gravity.

\begin{acknowledgement}
We want to thank our collaborators E. Okon, L. Ort\' iz, I. Pe\~na, D. Bedingham for their involvement and contribution towards this project. SKM is an International Research Fellow of Japan Society for the Promotion of Science. This research is partly funded by Grant-in-Aid to JSPS fellows (KAKENHI-PROJECT-15F15021). DS  acknowledges partial financial support from DGAPA-UNAM project IG100316 and  by CONACyT project 101712. 
\end{acknowledgement}
%
\addcontentsline{toc}{section}{Appendix}
%
%

%
%
%

\end{document}